\documentclass{mn2e}
\newcommand{\kms  }{\,\mathrm{km}\,{\mathrm{s}^{-1}}}

\newcommand{\La   }{\hbox{$\hbox{L\,$\alpha$}$}}
\newcommand{\ha   }{\hbox{$\hbox{H\,$\alpha$}$}}
\newcommand{\hb   }{\hbox{$\hbox{H\,$\beta $}$}}
\newcommand{\hc   }{\hbox{$\hbox{H\,$\gamma $}$}}
\newcommand{\hei  }{\hbox{$\hbox{He\,{\sc i}}$}}
\newcommand{\heia }{\hbox{$\hbox{He\,{\sc i 4471}}$}}
\newcommand{\heib }{\hbox{$\hbox{He\,{\sc i 4921}}$}}
\newcommand{\heic }{\hbox{$\hbox{He\,{\sc i 6678}}$}}
\newcommand{\heii }{\hbox{$\hbox{He\,{\sc ii}}$}}

\newcommand{\oia  }{\hbox{$\hbox{O\,{\sc i 7773}}$}}
\newcommand{\oib  }{\hbox{$\hbox{O\,{\sc i 8446}}$}}
\newcommand{\msun}{{\rm M}_\odot}

\title{GD 552: a cataclysmic variable with a brown dwarf companion?}
\author[Unda-Sanzana et al.]{
E. Unda-Sanzana$^1$, T. R. Marsh$^2$, B. T. G\"{a}nsicke$^2$, P. F. L. Maxted$^3$,
	\newauthor
        L. Morales-Rueda$^4$, V. S. Dhillon$^5$,  T. D. Thoroughgood$^5$, E. Tremou$^6$,
        \newauthor
        C. A. Watson$^5$, R. Hinojosa-Go\~ni$^1$\\
        $^1$ Instituto de Astronom\'ia,
        Universidad Cat\'olica del Norte, Antofagasta, Chile; eundas@almagesto.org; rhg002@ucn.cl\\
        $^2$ Dept. of Physics, University of Warwick, 
        Coventry, United Kingdom CV4 7AL; t.r.marsh@warwick.ac.uk; boris.gaensicke@warwick.ac.uk\\
        $^3$ Dept. of Physics and Astronomy, Keele University, 
        Keele, Staffordshire ST5 5BG, United Kingdom; pflm@astro.keele.ac.uk\\
        $^4$ Dept. of Astrophysics, University of Nijmegen, 
        6500 GL Nijmegen, The Netherlands; lmr@astro.ru.nl\\
        $^5$ Dept. of Physics and Astronomy, University of Sheffield, 
        Sheffield S3 7RH, United Kingdom; vik.dhillon@sheffield.ac.uk, C.Watson@sheffield.ac.uk\\
        $^6$ Dept. of Physics, Sect. of Astrophysics, Astronomy \&
        Mechanics, Univ. of Thessaloniki, 541 24 Thessaloniki, Greece; etremo@physics.auth.gr
}
\date{Accepted .
      Received ;
      in original form }
\pagerange{\pageref{firstpage}--\pageref{lastpage}}
\pubyear{2007}
\usepackage{graphicx}
\begin{document}
\maketitle
\label{firstpage}
\begin{abstract}
GD~552 is a high proper motion star with the strong, double-peaked
emission lines characteristic of the dwarf nova class of cataclysmic
variable star, and yet no outburst has been detected during the past 12 years
of monitoring. We present
spectroscopy taken with the aim of detecting emission from the mass
donor in this system. We fail to do so at a level which allows us to
rule out the presence of a near-main-sequence star donor. Given
GD~552's orbital period of $103$ minutes, this suggests that it is either
a system that has evolved through the $\sim 80$-minute
orbital period minimum of cataclysmic variable stars and now has a
brown dwarf mass donor, or that has formed with a brown dwarf donor in
the first place. This model explains the low observed orbital
velocity of the white dwarf and GD~552's low luminosity. It is also
consistent with the absence of outbursts from the system.
\end{abstract}
\section{Introduction}
\label{sec:introduction}
GD~552 is a blue, high proper motion star ($0.18''/$yr) discovered by Giclas et al. (1970). It
was first observed spectroscopically by Greenstein \& Giclas (1978), who found
that it is a cataclysmic variable star (CV) in which a white
dwarf accretes matter from a low-mass companion (often termed primary and
secondary star, respectively). GD~552's proper motion and position close to the
plane of the Galaxy (galactic latitude $4^\circ$), combined with its blue colour
all suggest that it is relatively close by. Greenstein \& Giclas (1978)
suggest a distance of $\sim 70\,$pc which gives a transverse motion which is
reasonable for  a member of the Galactic disk, and
combined with its magnitude $V = 16.5$, suggests $M_V \sim 12.5$, the equivalent of
a $0.6\,\msun$ white dwarf with a temperature of only $9000\,$K. The probable 
low luminosity of GD~552, which is of central importance to this paper, is backed
up by an absence of any observed outbursts, suggesting that it may very rarely or
never have outbursts (Cannizzo 1993).

The main spectroscopic study of GD~552 so far has been carried out by Hessman \&
Hopp (1990) (hereafter HH1990) who determined the orbital period of GD~552 of 
$102.7\,$min. They observed an extreme Balmer decrement (\ha:\hb = 6.2:1.0), 
indicative of a cool, optically thin disc, another indication of low luminosity
(Williams \& Shipman 1988).
HH1990 measured the white dwarf's projected orbital velocity to be $K_1 = 17\pm4
\kms$. This is a very low value suggestive of a low inclination system, since
for edge-on systems with orbital periods similar to GD~552 $K_1$ is typically
$\sim60-80 \kms$. However, the emission lines from the accretion disc 
suggest instead a moderately inclined system as they display clearly separated
double-peaks (which come from the outer disc, Horne \& Marsh 1986) with velocities of $\pm450 \kms$
which can be compared to typical peak velocities of $\sim600 \kms$ for edge-on
systems.
To solve this conundrum, HH1990 suggested that the
white dwarf in GD~552 is unusually massive -- close to the Chandrasekhar limit
in fact -- allowing a low orbital inclination ($\sim 20^\circ$) and therefore
the small $K_1$ value at the same time as large disc velocities. HH1990
were forced to their model because they assumed that the companion to the
white dwarf had to be a main-sequence star.  Since the companion fills its Roche
lobe, Roche geometry and the orbital period uniquely specify its density
($36.5\,{\rm g}\,{\rm cm}^{-3}$), which, if it is a main-sequence star, also
fixes its mass, which turns out to be $M_{2} \sim 0.13\,\msun$. This would cause a
much larger $K_1$ than observed, unless the inclination is low.

More recent theoretical work has revised HH1990's estimate a little to 
$M_{2} \sim 0.15\,\msun$ from GD~552's orbital period (Kolb \& Baraffe 1999), but the
problem is qualitatively unchanged. However, this work also suggests a very
different scenario. Cataclysmic variables at long orbital periods are thought to
evolve towards shorter periods until the companion becomes degenerate at a period near 70
minutes.  After this time, the mass donor increases in size as it loses mass and
the orbital period lengthens (e.g. Howell, Nelson \& Rappaport, 2001; Kolb 1993). Thus
although if GD~552 is approaching the period minimum its donor mass must be 
$M_{2} \sim 0.15\,\msun$, if it has already passed the minimum and is now evolving to longer
periods, it would be about a factor of four times less massive, and there would
be no need for HH1990's massive white dwarf, face-on model. Moreover,
such a system would have an extremely low mass transfer rate, consistent with
GD~552's lack of outbursts and probably low intrinsic luminosity. The
evolution just described is the standard explanation for the observed minimum
orbital periods of CVs, which however fails on two counts.  First, the observed
minimum around 80 min is distinctly longer than the theoretical value of 70 min
(Kolb \& Baraffe 1999; although Willems et al. (2005) propose that the period
minimum might be longer than expected because of extra angular momentum loss from
circumbinary disks), and, second, while we expect most systems to have passed
the period minimum there is just one single well-established example of such a
system known (SDSS 103533.03+055158.4, Littlefair et al. 2006)
although there are a few other good candidates (Mennickent et al. 2001; 
Littlefair, Dhillon \& Martin 2003; Patterson, Thorstensen \& Kemp 2005;
Araujo-Betancor et al. 2005; Southworth et al. 2006). 
Therefore it is of considerable interest to establish whether
GD~552 is a pre- or post-period-minimum system.

In this article we carry out a test to distinguish between the two models of GD 552. If
the pre-period-bounce model is correct, then, as we show in this paper, we should be able to detect features
from the M star in a low mass-transfer rate system such as GD~552. If, conversely,
the post-period-bounce model is correct, the donor will be extremely faint and it 
should not be detectable. In summary, our task is to set the strongest possible 
limit upon the presence of a hypothetical M dwarf such that we can say that we would 
have seen it had GD~552 been a pre-bounce system.

\section{Observations and data reduction}
\label{sec:observations}

We carried out observations of GD 552 in January and August 2001 on the island
of La~Palma in the Canary Islands (see Table \ref{tab:data} for
details). In January 2001, the 2.5-m Isaac Newton Telescope (INT) was
used in conjunction with the Intermediate Dispersion Spectrograph
(IDS) to acquire one dataset. Another dataset obtained in January 2001
used the 4.2-m William Herschel Telescope (WHT) in conjunction with
the double-beamed high-resolution ISIS spectrograph.  In August 2001
only WHT/ISIS was used. The weather was good during both runs, with 
no clouds and a typical seeing of 1 arcsecond.

An additional low-resolution spectrum was acquired on January 4,
2005, again with WHT/ISIS (see Table \ref{tab:data} for details).
This spectrum was acquired with a vertical slit in order
to obtain approximately correct relative fluxes. It was calibrated
using observations of Feige 34 (Oke 1990). The seeing on this night was around $2.5"$
and variable and so the flux calibration is only approximate. 
See Table \ref{tab:data} for details.
%Another
%single low-resolution spectrum was acquired on September 25, 2006
%using a similar setup (see Table \ref{tab:data} for details).

\begin{figure*}
\begin{center}
\includegraphics[width=5in,angle=270]{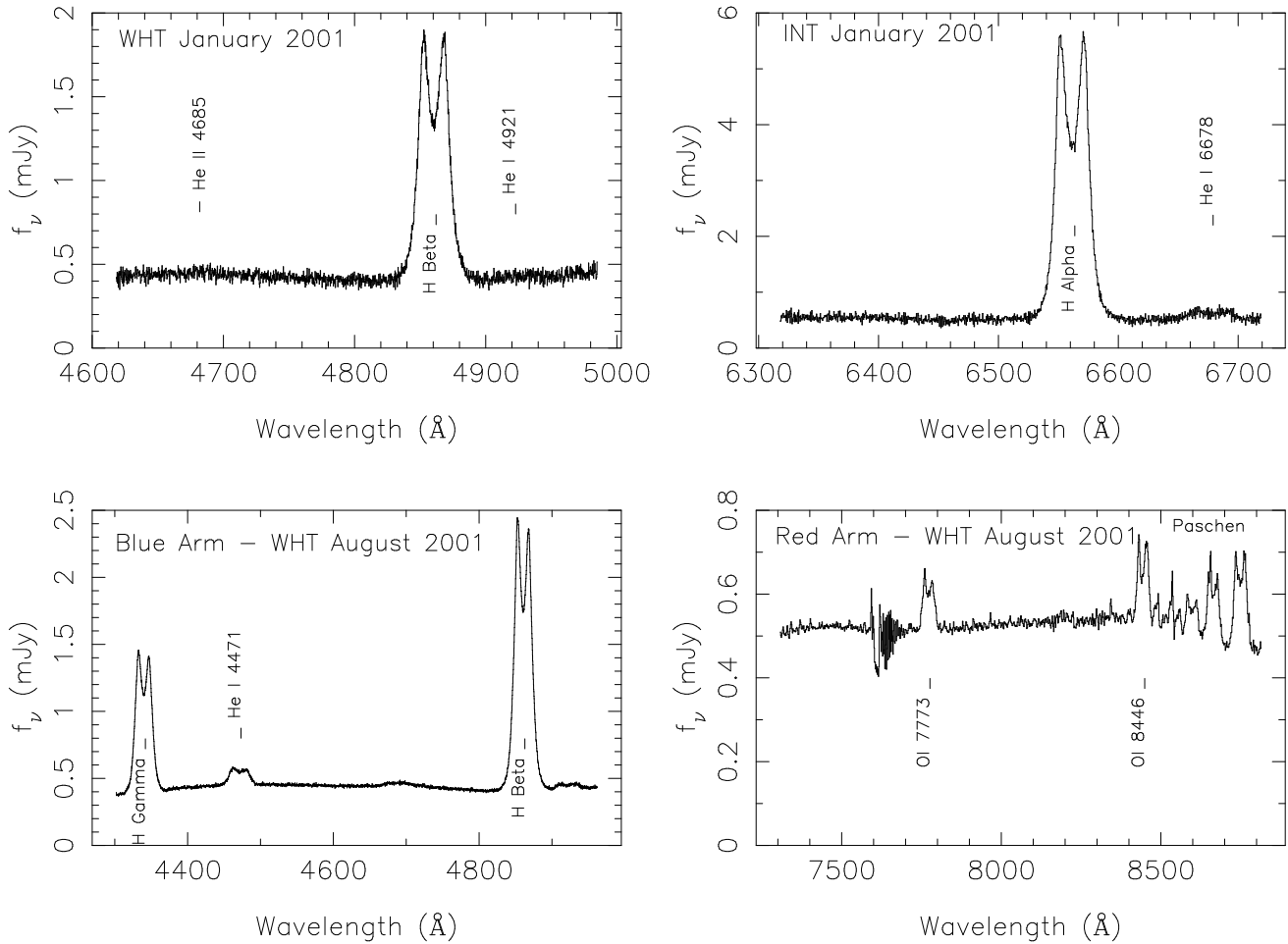}
\caption{\small Average spectra for the high-resolution datasets. A correction for 
telluric absorption has been made to the I-band spectra, although the very heavy
absorption at 7600~\AA\ could not be successfully removed.}
\label{fig:av}
\end{center}
\end{figure*}

\begin{figure}
\begin{center}
\includegraphics[width=4.8in,angle=270]{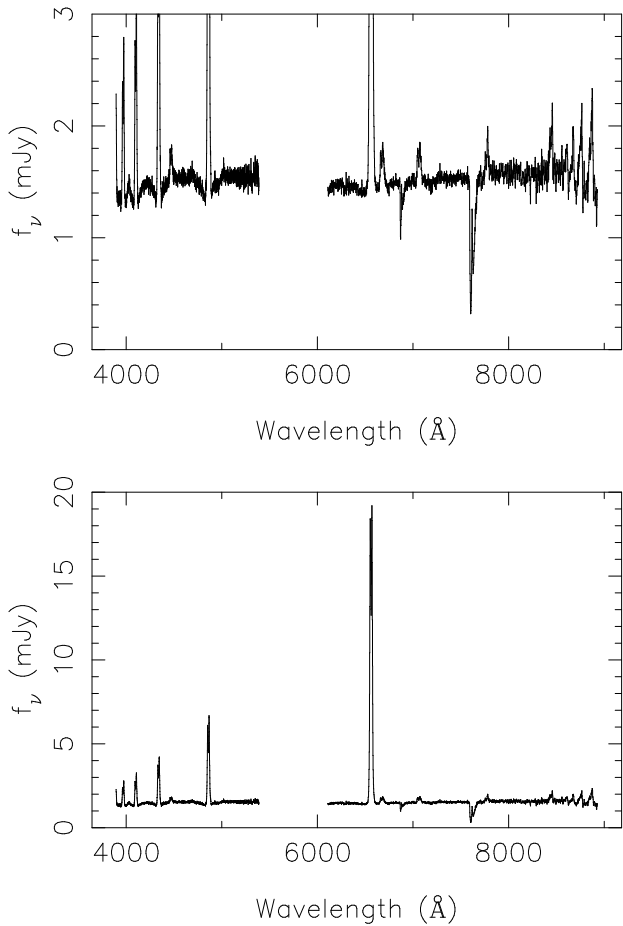}
\caption{\small Low resolution spectra acquired in January 2005. The upper
panel is a vertical enlargement of the lower one. Note the broad absorption wings around
the emission lines.}
\label{fig:lowresolutiondata2005}
\end{center}
\end{figure}

%\begin{figure}
%\begin{center}
%\includegraphics[width=2.4in,angle=270]{./Figure_Observations_Low_Resolution_2006.ps}
%\caption{\small Low resolution spectra acquired in September 2006. The flux scale
%is uncalibrated and has been normalised by the maximum value in the observed range.
%Note again the broad absorption wings around the emission lines.}
%\label{fig:lowresolutiondata2006}
%\end{center}
%\end{figure}

We flux calibrated the earlier spectra using observations of
the spectrophotometric standard HD19445 which we also used to remove 
the effect of telluric lines on the red
data (Bessell 1999). Flat-fields and comparison arc spectra were 
taken at regular intervals (every $\sim60\,$min).

The spectra were optimally extracted (Marsh 1989), with flat fields
interpolated from the many ones taken during the night. The wavelength 
scales for each science spectrum were obtained by interpolating the 
solutions of the nearest two arc spectra. For the
January 2001 datasets a comparison star was included in the slit.
In this case slit losses were corrected using the ratio of the 
spectra of the comparison star to a spectrum taken with a 
wide slit close to the zenith. For the August 2001 datasets no
comparison star was observed, so we simply normalised the continua of
these spectra.

Six images were acquired with the WHT on 25 September 2006 using a Harris
I-band filter. One Landolt standard (Mark A1) with very similar
airmass was observed to flux calibrate the images. 
Observations from the nearby Carlsberg Automatic 
Meridian Circle\footnote{http://www.ast.cam.ac.uk/$\sim$dwe/SRF/camc\_extinction.html} 
on the same night show that the night was mostly photometric 
with normal extinction.

Finally, 5.4h of filterless CCD photometry of GD552 was obtained on 29
July 2005 using a SI-502 $516\times516$ pixel camera on the 1.2m
telescope at Kryoneri Observatory. The data were reduced using the
pipeline described in G\"ansicke et al. (2004), where we used USNO-A2.0
1500-09346766 (located $\sim1\arcmin$ north-east of GD552, $B=16.8$,
$R=15.6$) as comparison star.

Ultraviolet STIS spectroscopy of GD552 was retrieved from the HST
archive. The system was observed with the far-ultraviolet (FUV) G140L
grating and the $52\arcsec\times0.2\arcsec$ aperture on 24 October
2002 for a total of $\sim4$\,h. The observation was split into 15
individual  exposures of 970\,s each. An additional three 1410\,s
near-ultraviolet spectra using the G230LB grating were obtained on 31
August 2002, again using the $52\arcsec\times0.2\arcsec$ aperture.

\begin{table*}
\caption{Summary of the WHT and INT spectroscopic data used in the analysis. In this table MD stands 
for 'mean dispersion'. FWHM is the full width at half maximum of the unblended arc lines.
T is the mean exposure
time per frame and DT is the average dead time between exposures. N
is the number of spectra collected per night per arm.}
\begin{tabular}{llccccccrrr}
\multicolumn{1}{l}{Telescope/}         &
\multicolumn{1}{l}{CCD/Grating}        &
\multicolumn{1}{c}{Date}               &
\multicolumn{1}{c}{Start - End}        &
\multicolumn{1}{c}{Orbits}             &
\multicolumn{1}{c}{$\lambda$ range}    &
\multicolumn{1}{c}{MD}                 &
\multicolumn{1}{c}{FWHM}               &
\multicolumn{1}{c}{T/DT}               &
\multicolumn{1}{c}{N}                  \\
\multicolumn{1}{l}{Instrument}         &
\multicolumn{1}{l}{}                   &
\multicolumn{1}{c}{}                   &
\multicolumn{1}{c}{(UT)}               &
\multicolumn{1}{c}{covered}            &
\multicolumn{1}{c}{(\AA)}              &
\multicolumn{1}{c}{(\AA\,pixel$^{-1}$)}&
\multicolumn{1}{c}{(\AA)}              &
\multicolumn{1}{c}{(s)}                &
\multicolumn{1}{c}{}                   \\
& & & & & & & &\\
% January 2001
INT/IDS &EEV10/R1200B&12/13 Jan 2001&19:48-22:06&1.343&6318-6719&0.39&0.78&300/30&26\\
WHT/ISIS&EEV12/H2400B&12/13 Jan 2001&19:52-21:13&0.788&4618-4985&0.21&0.42&300/12&32\\
\\
% August 2001
WHT/ISIS&EEV12/R1200B&13/14 Aug 2001&23:28-05:34&2.979&4301-4962&0.22&0.69&290/16&61\\
WHT/ISIS&TEK4/R316R  &13/14 Aug 2001&23:35-05:34&2.911&7306-8814&1.48&3.43&300/5 &63\\
\\
%Extra data
WHT/ISIS&EEV12/R1200B&04/05 Jan 2005&19:50-19:58&0.081&3900-5390&0.88&1.8 &500/0&1\\
WHT/ISIS&TEK4/R316R  &04/05 Jan 2005&19:50-19:58&0.081&6110-8930&1.65&3.2 &500/0&1\\
%WHT/ISIS&EEV12/R1200B&25/26 Sep 2006&23:28-05:34&0.097&3644-5265&0.88&1.8 &600/0&1\\
\end{tabular}
\label{tab:data}
\end{table*}

\section{Analysis}

\subsection{Average profiles and trailed spectra}
\label{sec:average_profiles_and_trailed_spectra}

Figure \ref{fig:av} shows average spectra for our data. Figure
\ref{fig:lowresolutiondata2005} 
%and \ref{fig:lowresolutiondata2006}
shows the available low resolution spectra. We observe
double peaked profiles in all the detected lines: \hei\ 4471.68~\AA\
(hereafter \heia), \heii\ 4685.750 \AA\ (hereafter \heii), \hei\
4921.93 \AA\ (hereafter \heib), \hei\ 6678.15 \AA\ (hereafter \heic),
\oia, \oib\ and the Paschen and Balmer series. In the general theory of
CVs, the double peaked profiles are consistent with disc emission.

\begin{figure*}
\begin{center}
\includegraphics[width=2.8in,angle=270]{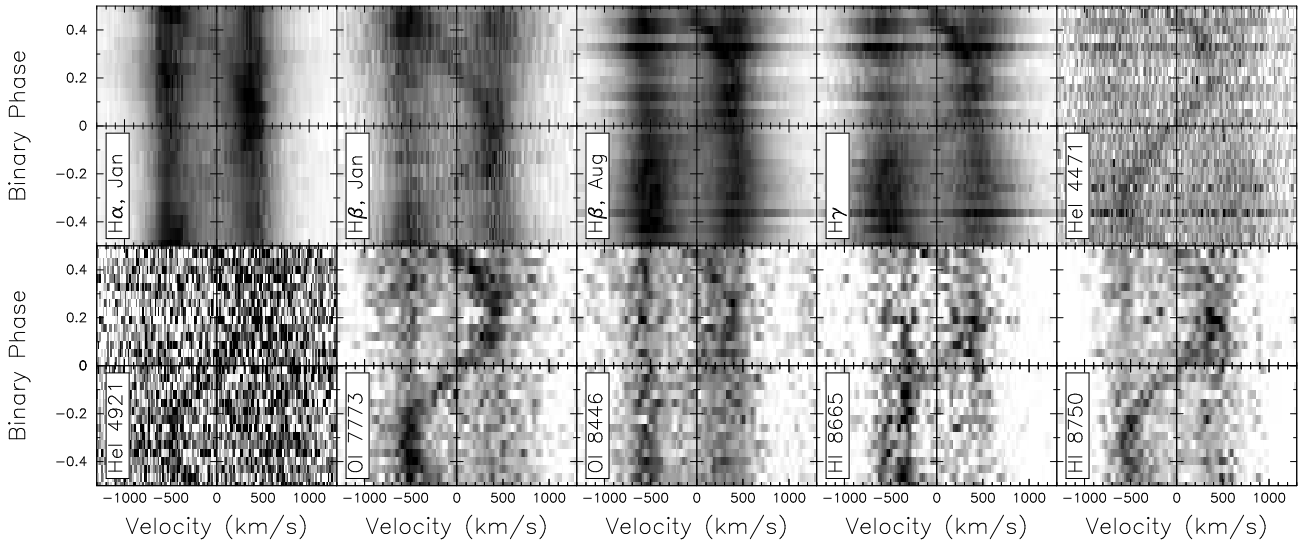}
\caption{\small Phase binned data. The fluxes in this figure have been normalised by the 
maximum value the flux reaches for each wavelength. We set the colour scale 
to white for 0 flux and to black for 85\% of this normalised flux, with the
exception of line \heib for which black corresponds to 
40\% of the normalised flux. 
%Please see Section \ref{sec:doppler_tomography} for
%an explanation of the different phasing in the datasets.
}
\label{fig:tr}
\end{center}
\end{figure*}

A second order polynomial was fitted to the continuum of each dataset
and then the data were divided by these fits. The normalised continua
were subtracted, and the datasets were binned in 20 (Jan) or 30 (Aug)
orbital phase bins before plotting Figure \ref{fig:tr}.
%The features
%we detected show emission lines of very different strengths from one
%another, and thus it is not possible to find a suitable common scale
%to appreciate detail in all of them. Instead, when needed we divided
%the dataset producing one panel for each wavelength and then we
%normalised the gray scale from the continuum level to a maximum
%intensity in each panel.
These trailed spectra display double peaked
emission following a sinusoidal motion with orbital
phase. We interpret this as the rotating accretion disk, approximately
tracking the emission of the white dwarf it surrounds. The other
remarkable feature in these spectra is a higher-amplitude sinusoid
also varying with orbital phase, but shifted with respect to the
assumed zero phase. It can be seen very clearly, for instance, in
\oia\ (Figure \ref{fig:tr}). This emission is produced in the
stream/disc impact region, which is usually termed the bright spot.
Note that the phasing of this sinusoidal is different for the January
and August datasets. This is due to the uncertainty in the zero phase
of the ephemeris used to fold the data (HH1990's).

Using the average of the WHT spectra acquired in 
January 2005, we measured the Balmer decrement defined as the ratio of line
fluxes $\ha:\hb:\hc$. We did this by integrating
the flux from each profile after subtracting a low order fit to the
continua. We computed $\ha:\hb:\hc=2.0:1.0:0.56$.  
Our values for the Balmer decrement differ noticeably from HH1990's
($\ha:\hb:\hc=6.2:1.0:0.26$), being rather closer to Greenstein \& Giclas (1978)
($\ha:\hb:\hc=2.2:1.0:0.5$). Qualitatively, though, our conclusion is 
the same: the Balmer decrements suggest that the accretion disk is 
optically thin and cool (see, for example, Williams \& Shipman, 1988). This 
conclusion is strengthened by the detection of \oia\ which, according 
to Friend et al. (1988) is a good indicator of the state of the disc, 
its emission indicating an optically thin accretion disc.

%This is discussed in more detail
%in Section \ref{sec:doppler_tomography}.

\subsection{The primary star}
\label{sec:the_primary_star}

HH1990 measured a value of $17\pm4 \kms$ for $K_{1}$. We tried to
check and refine this value by using Schneider \& Young (1980)'s method.
We convolved our data with a difference of Gaussians equidistant from a 
candidate line centre, and define the velocity as the 
displacement of the double Gaussians such that the convolution equals
zero. By increasing the separation of the
Gaussians we gather information from the wings of the profile, which is 
produced in regions of the disc closer to its centre (Horne \& Marsh 1986).
We assume that the motion of these regions tracks the motion of the accreting 
object, so the further into the wings the measurement is made, the better the 
estimate of the motion of the white dwarf. The process does not 
continue indefinitely because there is a maximum velocity in the line wings,
which corresponds to the Keplerian motion of the innermost ring of gas just
before settling onto the white dwarf. However, in practice, before reaching
this limit the reliability of the calculation is constrained by the noise 
present at the continuum level.

The motion of each disc region is calculated by fitting the orbital
solution:
\begin{equation} 
\label{eq:vsine}
V = \gamma - K \sin\left(\frac{2\pi(t-t_{0})}{P}\right)
\end{equation}
to the measurements. $\gamma$ is the systemic velocity, $K$ is the velocity 
of the source of the emission, $t_{0}$ is the time of superior conjunction
of the emission-line source, and P is the orbital period of the system.
Further, we avail ourselves of a diagnostic diagram (Shafter, Szkody \& Thorstensen
1986) to decide when the calculation is becoming dominated by noise as the
separation of Gaussians ($a$) increases. This should reveal itself as a sharp
rising of the statistics $\sigma_{K}/K$. At the same time, we also expect to see
convergence of the calculated parameters ($K$, $\gamma$) and a phasing appropriate
to the white dwarf.

\begin{figure}
\begin{center}
\includegraphics[width=3.7in,angle=270]{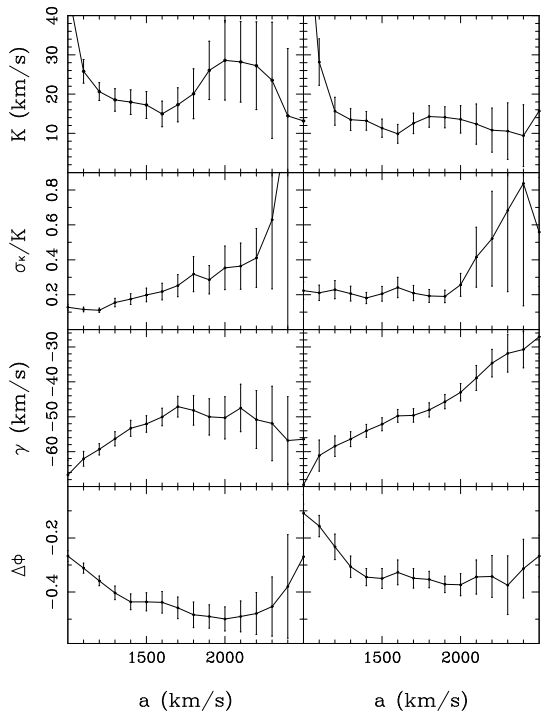}
\caption{\small Diagnostic diagram for \ha\ (left) and \hb\ (right), using INT January 
2001 and WHT August 2001 data respectively.}
\label{fig:dd}
\end{center}
\end{figure}

In our diagrams (Fig.~\ref{fig:dd}) we do not put a strong constraint on the value of
convergence for the phasing (bottom panels) because the
ephemeris of this system has a large uncertainty and thus its orientation
is arbitrary for the date of our observation.
For $a$ larger than $2500 \kms$ the noise dominates both diagrams so this is the upper limit
for the abscissa in our plots. 
The \hb\ diagram displays a reasonable degree of convergence
when $a$ approaches $2000 \kms$. On reaching that value the noise takes off.
On the other hand, for large values of $a$ in \ha\ we are picking data
contaminated by the wings of \heic\ and thus we do not expect
a clear convergence in this diagram, although the noise seems to dominate
once $a$ reaches $2200 \kms$. Eventually, as it is not obvious which value
to favour in \ha\, we take as $K_{1}$
the value of $K$ for $a = 1900 \kms$
in \hb, right before the noise begins to dominate. This is $14.1\pm2.7 \kms$
consistent with HH1990. The uncertainties were estimated
by means of the bootstrap method, repeating the calculation for 10000 bootstrap
samples (Diaconis \& Efron 1983).

The high resolution B-band spectrum of Fig.~\ref{fig:av} shows signs of broad
absorption wings flanking the Balmer lines. These are often seen in low
luminosity dwarf novae (e.g. Szkody et al. 2007) and originate in
the photosphere of the white dwarf. Low resolution spectra taken in January
2005 and September 2006 confirm the presence of the white dwarf
photospheric lines (Fig.~\ref{fig:lowresolutiondata2005}). The white dwarf is also revealed in
HST/STIS ultraviolet spectra through the a very broad \La\ feature
extending to 1600A (Fig.~6). This feature, which results from \La\
from neutral hydrogen plus quasi-molecular features at 1400 and 1600 \AA\ from
$H_{2}^{+}$ and $H_{2}$, is a hallmark of cool white dwarfs.

%The left-hand panels in Fig.~\ref{fig:av} show signs of broad
%absorption wings around the Balmer lines, indicative of the white
%dwarf. If true, modelling the white dwarf spectrum can provide a
%constraint on the distance to GD~552, which is crucial to our attempt 
%to detect the mass donor. For the purpose of spectral modelling high-resolution
%data are not appropriate because of the associated problems in the flux
%calibration and the small wavelength coverage. We therefore decided to
%obtain additional low-resolution spectra with ISIS on the
%WHT (Figs.~\ref{fig:lowresolutiondata2005} and \ref{fig:lowresolutiondata2006})
%which clearly confirms the presence of broad
%white dwarf absorption lines at H$\beta$ to H$\delta$. White dwarfs can dominate
%low mass transfer rate systems in the ultraviolet and we therefore
%obtained STIS spectra of GD552 from the HST archive.

%The HST/STIS ultraviolet spectrum (Fig.~\ref{fig:stis}, left panel)
%reveals an extremely broad L$\alpha$ absorption line extending to
%$\sim1600$\,\AA, the hallmark of a relatively cool white dwarf. The
%extreme width of the L$\alpha$ line is caused by the superposition of
%L$\alpha$ from neutral hydrogen plus its quasimolecular 1400\,\AA\
%H$_2^+$ and 1600\,\AA\ H$_2$ satellite lines. 

\begin{figure*}
\includegraphics[angle=270,width=\columnwidth]{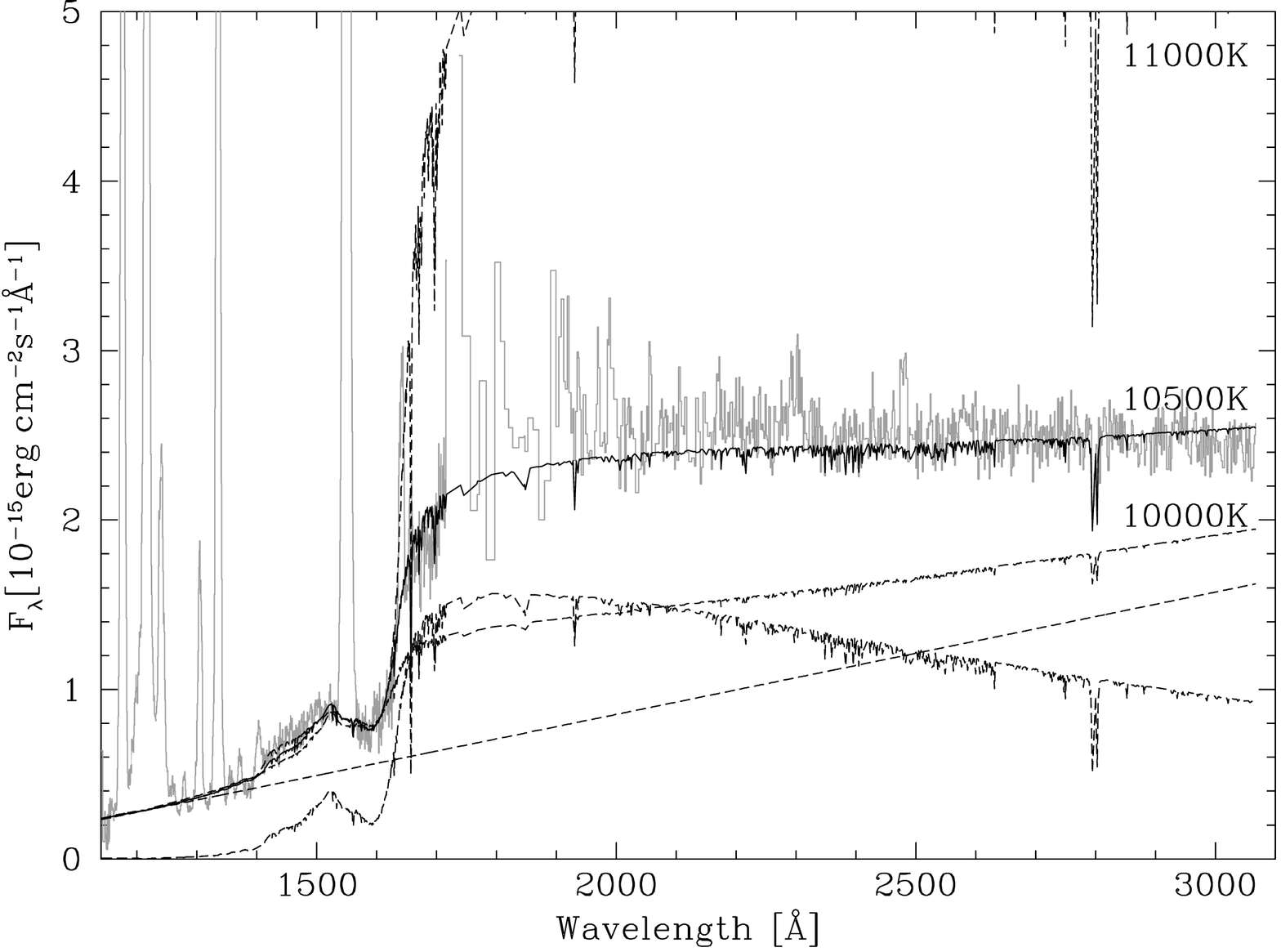}
\hfill
\includegraphics[angle=270,width=\columnwidth]{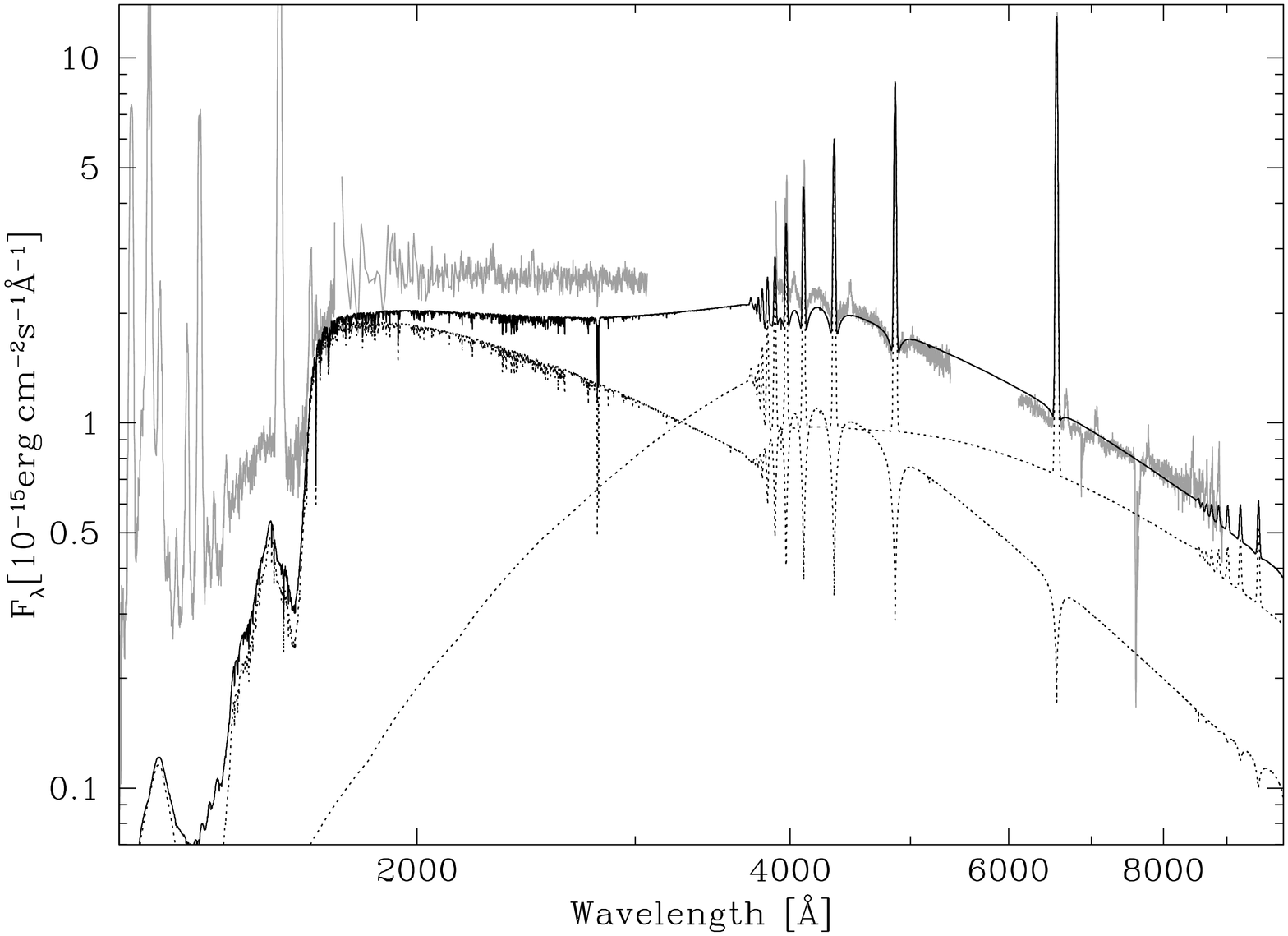}
\caption{\small Left: The archival STIS ultraviolet spectrum of GD552,
  plotted in gray. The solid line shows a two-component fit consisting
  of a linear slope plus a white dwarf model spectrum
  ($T_\mathrm{eff}=10\,500$\,K, $\log g=7.5$), both components are
  plotted individually as dashed lines. Shown as dotted line are the
  same two-component models, but for $T_\mathrm{eff}=11\,000$\,K (top
  curve) and 10\,000\,K (bottom curve). Right: The STIS ultraviolet
  spectrum along with our WHT low-resolution spectrum, plotted in
  gray. Plotted as solid black line is the sum of the white dwarf
  model as shown in the left panel plus the spectrum of an
  isothermal/isobaric hydrogen slab. Both model components are
  plotted individually as dotted lines. }
\label{fig:stis}
\end{figure*}

For the purpose of the spectral analysis, we have computed a grid of
white dwarf models covering effective temperatures in the range
8000\,K--15\,000\,K and surface gravities of $\log g=7.5$, 8.0, and
8.5 using the stellar atmosphere and spectral synthesis codes
TLUSTY/SYNSPEC (Hubeny \& Lanz 1995). 
Metal abundances were set to 0.1
times their solar values, as suggested by the weakness of the
Mg$\lambda\lambda$2796,2803 resonance doublet.
The assumed value for the metal abundance is within the typical range for CVs
(Sion \& Szkody 2005).
A first exploratory
fit revealed that none of the white dwarf model spectra could
reproduce the STIS data alone, as the flux of the model spectra
rapidly drops to zero shortward of $\sim1300$\,\AA, in contrast to the
observed spectrum that has a substantial non-zero flux down to
1150\,\AA. Moreover, it turns out that the 15 individual
far-ultraviolet spectra show substantial flux variability, supporting
the presence of a second (variable) component. The ultraviolet flux
variability shows a clear dependence on wavelength, and we have fitted
the continuum slope of the spectrum with a linear slope,
$F_\lambda=7.22\times10^{-19}\lambda-5.9\times10^{-16}\,\mathrm{erg\,cm^{-2}s^{-1}\AA^{-1}}$.
Qualitatively, such a slope is what could be expected from either an
optically thin emission region (where absorption is dominated by
bound-free absorption) or an optically thick relatively cool
quasi-blackbody component (where the far-ultraviolet emission is on
the Wien part of the blackbody spectrum).  We then explored
two-component fits to the STIS data consisting of this linear slope
plus a white dwarf model, with the contributions of the slope and the
white dwarf being normalised to match the observed continuum below
1300\,\AA\ and in the range 1580--1620\,\AA, respectively. The best
fit for $\log g=7.5$ is shown in Fig.\,\ref{fig:stis} (left
panel). This simple model provides a surprisingly good match to the
STIS spectrum over the entire ultraviolet range. However, an
extrapolation of this simple linear slope over a broader wavelength
range, i.e. into the optical, is not warranted, as the exact nature of
this emission component is not known.

The parameters determined from
the fit are the white dwarf effective temperature $T_\mathrm{eff}$,
and the flux scale factor between the synthetic and the observed
spectrum. Assuming a white dwarf radius, the flux scaling factor can
be used to calculate the distance to the system. As the white dwarf
mass is unconstrained by this analysis, we ran the fit for $\log
g=7.5$, 8.0, and 8.5, corresponding to white dwarf masses and radii of
(0.33\,$\msun$, $1.15\times10^9$\,cm), (0.57\,$\msun$,
$8.65\times10^8$\,cm), and (0.9\,$\msun$, $6.15\times10^8$\,cm),
respectively (where we have assumed a Hamada-Salpeter (1961)
Carbon-Oxygen core mass-radius relation). The best fit parameters for
$\log g=7.5$, 8.0, and 8.5 are (10\,500\,K, 125\,pc), (10\,900\,K,
105\,pc), and (11\,300\,K and 85\,pc), respectively.  Because of the
very strong temperature dependence of the 1600\,\AA\ H$_2^+$ the STIS
spectrum constrains the white dwarf temperature within a very narrow
range. 

Fig.\,\ref{fig:stis} (right panel) shows that the white dwarf
model obtained from the fit to the STIS data extended into the optical
range clearly underpredicts the observed flux by $\sim50$\%. The
presence of double-peaked Balmer emission lines clearly reveals the
presence of an accretion disc in the system that is contributing light
in the optical. We model the accretion disc by an isobaric/isothermal
hydrogen slab, which has three free parameters: the disc temperature
$T_\mathrm{d}$, the column density $\Sigma$, and the flux scaling
factor (see G\"ansicke et al. 1997, 1999 for details). A disc
temperature of $T_\mathrm{d}=5800$\,K and a column density of
$\Sigma=2.3\times10^{-2}\mathrm{g\,cm^{-2}}$, scaled to the observed
H$\alpha$ emission line flux provides an adequate model of the optical
spectrum. The emission line flux ratios of the model are
$\ha:\hb:\hc=2.2:1.0:0.56$, fairly close to the observed values
(Sect.\,\ref{sec:average_profiles_and_trailed_spectra}). For a
distance of $\sim100$\,pc, the area of hydrogen slab implied by the
flux scaling factor is consistent with the size of the Roche lobe of
the white dwarf. The disc parameters found here are very similar to
those found in a number of other studies of quiescent accretion discs
in CVs (e.g. Williams 1980, Marsh 1987, Lin et al. 1988, Rodriguez-Gil et
al. 2005). We refrained from more detailed modelling of the optical
data as the spectrum was obtained under non-photometric conditions,
and the flux calibration is subject to some uncertainty. Nevertheless the temperature derived from 
the UV data is consistent with our detection of the white dwarf's photosphere
in our optical spectra.

\begin{figure}
\begin{center}
\includegraphics[angle=270,width=\columnwidth]{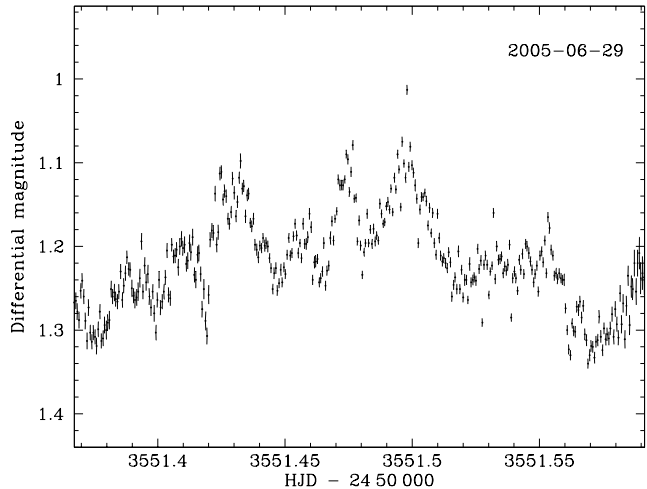}
\caption{\small Kryoneri CCD photometry of GD 552.}
\label{fig:photometry}
\end{center}
\end{figure}

The recent discovery of several non-radially pulsating white dwarfs in
CVs (e.g. van Zyl et al. 2004, Warner \& Woudt 2004, Araujo-Betancor et
al. 2005, Patterson et al. 2005, Warner \& Woudt 2005 and
G\"ansicke 2006) motivated the short
photometric time series observation of GD552 carried out at Kryoneri
observatory. Visual inspection of the light curve
(Fig.\,\ref{fig:photometry}) reveals substantial variability on time
scales of tens of minutes, plus a modulation over $\sim4.5$\,h. A
Lomb-Scargle period analysis (Lomb 1976, Scargle 1982) shows no substantial power at
periods shorter than $\sim15$\,min, where non-radial pulsations would
be expected to show up. The conclusions from this admittedly limited
set of data is that the white dwarf in GD 552 is probably too cool to
drive ZZ Ceti pulsations in its envelope and the varibility seen in the light
curve is probably due to flickering in the accretion disc.

\subsection{The secondary star}
\label{sec:the_secondary_star}

The presence of a secondary star (assumed to be an M dwarf since we are testing
for the presence of a star of mass $0.15\,\msun$) is not obvious either in the
average or in the trailed spectra. 

%Even if we had detected emission from the donor through Doppler tomography, it
%would not have implied the presence of an M dwarf. 

Our main aim was to 
detect photospheric absorption lines, especially the NaI doublet near 8200 \AA\ 
which is strong in late M dwarfs. To do so we compared our GD~552 spectra with
a set of M dwarf template spectra kindly provided by Kelle Cruz (personal
communication) following the classification by Kirkpatrick, Henry \& McCarthy
(1991). The stars used are listed in Table \ref{tab:m_templates}.

\begin{figure*}
\begin{center}
\includegraphics[width=2.7in,angle=270]{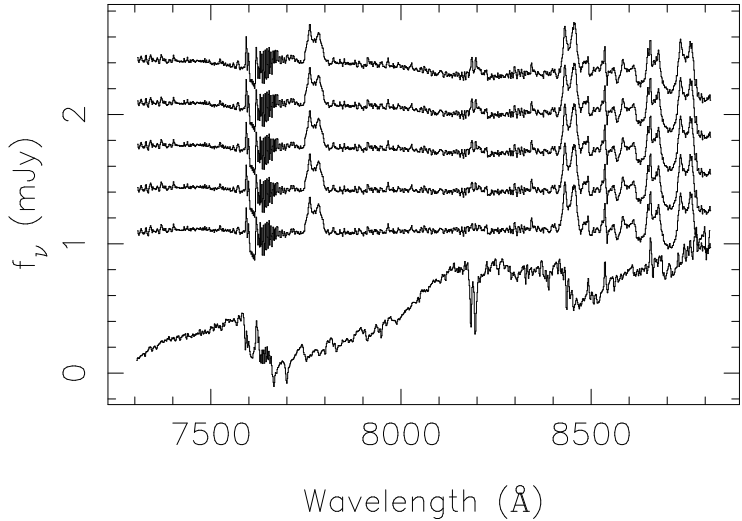}
\caption{\small From bottom to top: normalised template star (Gl 65A); normalised GD 552; and the four following curves
are normalised GD 552 minus 5\%, 10\%, 15\% and 20\% of the normalised template respectively.
When 10\% of Gl 65A is subtracted from GD~552 (third curve, top to bottom), the presence of
M star features is easily seen from the growing bump at 8200\AA\ due to the Na I doublet. The vertical
scale in the figure includes an offset of 0.5 between each plot and thus it is arbitrary.}
\label{fig:gd552minus_gl65a}
\end{center}
\end{figure*}

The Na I doublet near 8200 \AA\ and nearby molecular bands are the fingerprints
of M dwarfs. As we were not able to recognise any such features in the GD~552
data, we looked for a constraint upon its presence given that its features are
not detectable. First we normalised the spectra of GD~552 and of each template
in Table \ref{tab:m_templates}. We did this by dividing by a constant fit to the
continuum in the range 8100-8400 \AA, excluding the $\sim8200$ \AA\ doublet.
Then we subtracted each normalised template from the GD~552 data in 5\% steps.
In Figure~\ref{fig:gd552minus_gl65a} we show an example of this procedure.
GD~552 is at the bottom of the plot. Each profile above GD~552 increases the
amount of template subtracted from the data by 5\%. Consistently, we found that
for all the templates the presence of M star features is obvious once we reach
10\%.

\begin{table}
\caption{List of M star templates used for estimating
maximum contribution of an M star to the GD~552 profile.}
\begin{center}
\begin{tabular}{ll}
      &           \\
Spectral Type & Star \\
      &           \\
M0.5  & Gl 720A   \\
M1    & Gl 229    \\
M1.5  & Gl 205    \\
M2.5  & Gl 250B   \\
M3    & Gl 752A   \\
M3.5  & Gl 273    \\
M4    & Gl 213    \\
M4.5  & Gl 83.1   \\
M5.5  & Gl 65A    \\
M6    & Gl 406    \\
M6.5  & G 51-15   \\
M7    & VB 8      \\
M8    & VB 10     \\
M9    & LHS 2065  \\
\end{tabular}
\end{center}
\label{tab:m_templates}
\end{table}

\subsection{System parameters}
\label{sec:system_parameters}

In section \ref{sec:the_secondary_star} we placed an upper limit of 10\% upon
the contribution of an M star to the spectrum of GD~552 in the range 8100 to
8400\AA. We now use this restriction to estimate the mass of this hypothetical
star. This will allow us to compare directly to HH1990's $M_{2}$ value,
and thus to test their model.

\begin{figure}
\begin{center}
\includegraphics[width=2in,angle=0.0001]{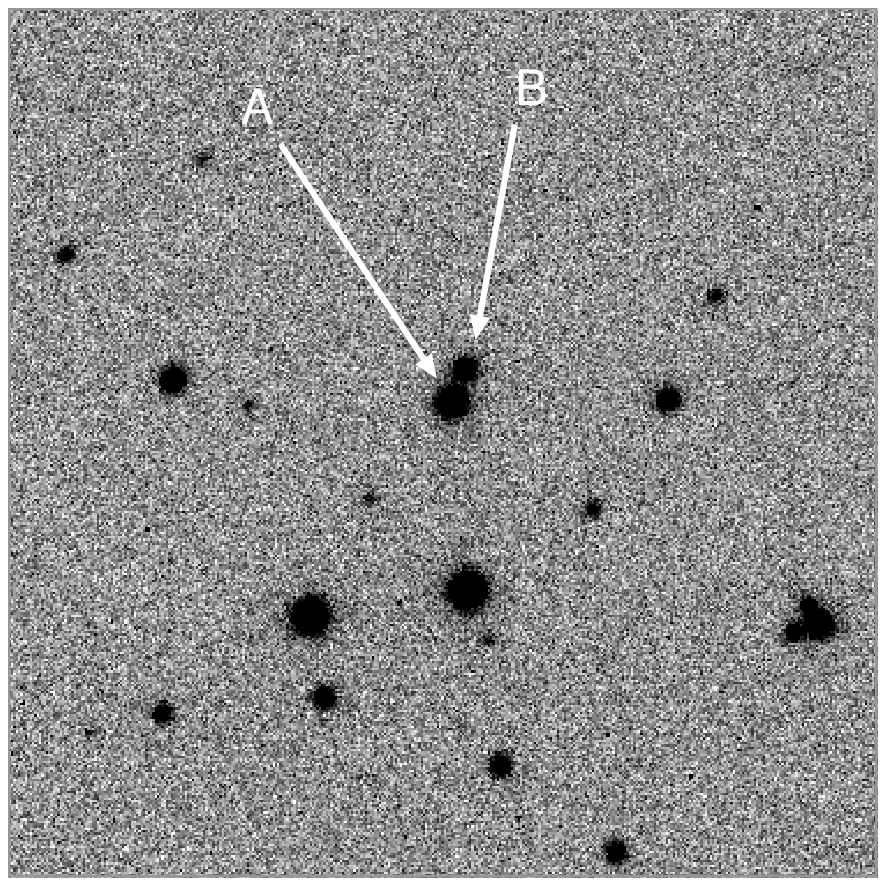}
\caption{\small GD~552 (marked with an A) and a field star (marked with a B) in
September 2006 (image scale is $\sim 1'$ by $1'$). The field star was in front of the system in previous years
and made impossible to estimate GD~552's magnitude with any certainty. The high proper motion
of GD 552 results in it and the field star now being resolved. Using our data 
we measured $I=16.3$ for GD~552 and $I=17.6$ for the field star.}
\label{fig:field}
\end{center}
\end{figure}

\begin{figure}
\begin{center}
\includegraphics[width=4in,angle=270]{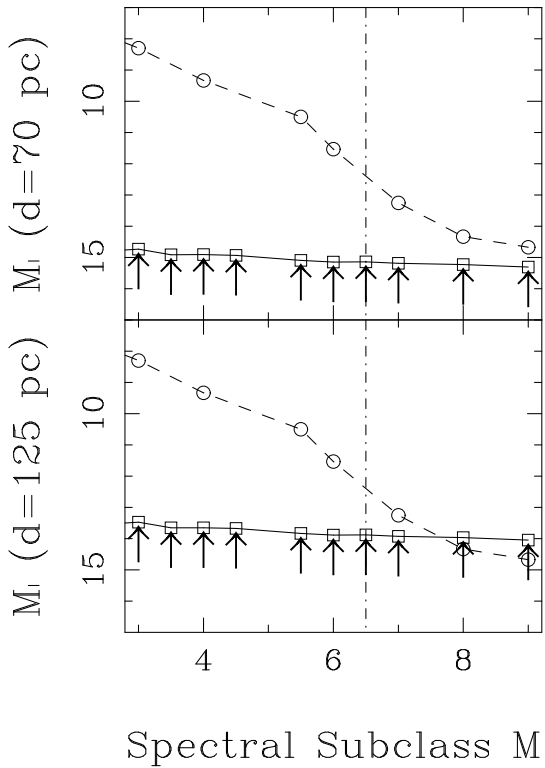}
\caption{\small Comparison of the upper limit on $M_{I}$ derived from our
I-band spectra for the mass donor in GD~552 (solid line, open squares plus arrows,
to indicate the upper-limit nature of the point)
and $M_{I}$ for young M dwarfs of the same spectral type (dashed line, open circles, Leggett 1992).
The vertical line at M6.5 marks the limit
set by Kirkpatrick \& McCarthy (1994) for a hydrogen burning star. The pannels assume
different distances to the system. The upper one uses the old Greenstein \& Giclas (1978) value
while the lower panel uses the distance estimated in this paper. }
\label{fig:magnitudes}
\end{center}
\end{figure}

We first need to convert our constraint into one upon the $I$-band magnitude
of the M star. We start from the relation between magnitudes and 
fluxes, including a colour correction to make our data match the profile of 
the I band:
\begin{equation} 
\label{eq:mi10mstar}
%m_{I, 10\%MS} = m_{I, GD~552} - 2.5 \log\left(\frac{\int^{\lambda_{f}}_{\lambda_{i}} \epsilon_{I}(\lambda) \lambda f_{\lambda, 10\%MS} d\lambda}{\int^{\lambda_{f}}_{\lambda_{i}} \epsilon_{I}(\lambda) \lambda f_{\lambda, GD~552} d\lambda}\right)
m_{I, 10\%MS} = m_{I, GD~552} - 2.5 \log\left(\frac{\int^{\lambda_{f}}_{\lambda_{i}} \epsilon_{I} \lambda f_{\lambda, 10\%MS} d\lambda}{\int^{\lambda_{f}}_{\lambda_{i}} \epsilon_{I} \lambda f_{\lambda, GD~552} d\lambda}\right)
\end{equation}
In this equation, $10\%MS$ stands for 10 percent of the template M star;
$\epsilon_{I}$ is the transmission for the I Band; $\lambda_{i}$ and
$\lambda_{f}$ are the limits of the observed range in wavelength. $f_{\lambda,
10\%MS}$ is 10\% of the normalised flux density of the M star, obtained by dividing by a
constant fit to the 8100-8400 \AA\ region (the Na I doublet at 8200 \AA\ 
excluded) and multiplied by 0.10.  Similarly, $f_{\lambda, GD~552}$ is the flux
density of GD~552 calculated by dividing by a constant fit to its continuum.
%Using the data acquired in September 2006 we measured $Harris-I = 16.3$. In
%what follows we assume the difference between I and Harris-I to be negligible,
%so we use $I = 16.3$. 
A field star in front of GD 552 made difficult to do this in previous years. 
By September 2006, however, the interloper had moved enough to make this calculation
straightforward (see Figure~\ref{fig:field}). We measured $m_{I,GD 552}=16.3$ from our data.
Then we use equation~\ref{eq:mi10mstar} with the M star templates listed
in Table~\ref{tab:m_templates}, obtaining $m_{I, 10\%MS}$ as a (mild) function of
the assumed spectral type (hereafter this will be referred to simply as
$m_{I}$). In Figure~\ref{fig:magnitudes} we compare these values after conversion
to absolute magnitude $M_I$ for distances of $70$ and $125\,$pc with 
the absolute magnitudes of young M dwarfs of the same spectral type from Leggett (1992).
In doing this we are assuming solar-like metallicities.

%(It must be noted that our data did not cover a small
%part of the blue wing of the I band, which affects the calculation according to
%Equation~\ref{eq:mi10mstar}, however tests adding extrapolated points to 
%estimate the difference with our result show the effect to be negligible.)

The vertical line in Figure~\ref{fig:magnitudes} marks the spectral type for a hydrogen burning star with
the lowest possible mass ($\sim 0.08\,\msun$, Kirkpatrick \& McCarthy 1994). The difference
between the top and bottom panels is the distance assumed in each case. In the top panel
we used the value $d \sim 70\,{\rm pc}$ from Greenstein \& Giclas (1978); it is clear from it
that the absolute magnitudes allowed by our data are inconsistent with the assumption
that the mass donor in GD~552 is a main-sequence star.
This implies that the mass of the companion star $M_{2} < 0.08 M_{\sun}$, ruling out
near main-sequence models such as that of HH1990. Our conclusion remains
unaltered even if GD~552 is at the maximum distance of 125~pc calculated for a $0.33 \msun$ white dwarf
(Section \ref{sec:the_primary_star}) which is lower than any white dwarf mass measured in any CV.

\section{Discussion}

Our model of GD~552 is radically different from that of HH1990. Where
they have a very high mass white dwarf in an almost face-on system, with an M
dwarf donor of mass $\sim 0.15\,\msun$, our failure to detect the mass donor
requires that it is a brown dwarf with $M_2 < 0.08\,\msun$. This makes GD~552 an
excellent candidate for a post-period-bounce system with an age of about 7 Gyr (Politano
et al. 1998). Our explanation of the low $K_1$ is therefore simply a case of extreme mass ratio,
and we do not require a particularly low orbital inclination. If the secondary is 
low mass, there is no need for the system to be face on to get
$K_{1} = 17.4 \kms$, and therefore neither is there any need for the white dwarf
to have a high mass to produce the double peaked profiles. 
For example, the observed value of $K_1$ can be matched with a white dwarf of 
mass $M_{1} = 0.6 \msun$ in orbit with a companion of mass $M_{2} = 0.03 \msun$ 
with a moderate inclination of $i \sim 60^{\circ}$.

Patterson et al. (2005) suggested that GD~552 is a likely ``period
bouncer'' (post period minimum CV) because of its lack of outbursts and small $K_{1}$. We have now
given support to their suggestion. However, HH1990's original model \emph{could} have
been correct and the search for the donor was a crucial step in ruling out
HH1990's hypothesis. Mass donors are easily visible in other systems with hotter
white dwarfs and with orbital periods similar to GD~552
(e.g. HT Cas, $P_{orb} = 106.1 min$, M5-M6V secondary spectral type, Marsh 1990; 
Z Cha, $P_{orb} = 107.3 min$, M5.5V secondary spectral type, Wade \& Horne 1988;
compared to $P_{orb} = 102.73 min$ for GD 552).
Our search for the mass donor was thus a realistic and critical test for the
suitability of HH1990's model.

There is a key point about our analysis that is worth emphasizing: we have
managed to derive particularly strong constraints because of GD~552's
relatively long orbital period which maximises the difference between the pre-
and post-period-bounce systems. In particular it means that the mass donor in
the pre-bounce model is relatively easy to detect, so that failure to detect it
is a clear indication of the post-bounce alternative. We believe that this
makes GD~552 one of the most secure post-period-minimum CV known (RE~J1255+266
may be even better if its relatively long period can be confirmed, see Patterson et
al. 2005). The method we employ is indirect, but necessarily so, since if the donor is a
brown dwarf it is likely to contribute far less light than the upper
limit we have derived.

It can be argued that there is also the possibility that GD552 is not a
post-bounce system at all, even if its components are a WD with a brown
dwarf donor. If GD552's progenitor was a binary composed of a main
sequence star and a brown dwarf, instead of a double main sequence binary,
it would have evolved into the CV we see today without the need of going
through the minimum period. We would not be able to distinguish between
these two systems at all. Politano (2004) found that CVs 
that evolve directly to a short orbital period are expected to be rare because 
the progenitor systems containing a solar-type star and a brown dwarf are 
known to be rare (``the brown-dwarf desert'').

%According to population synthesis
%calculations by Politano (2004), most zero age CVs with brown dwarf
%companions show orbital periods between 46 and 78 minutes with only a
%small number showing periods larger than 78 minutes. Moreover, their
%calculations predict that a CV + brown dwarf with an orbital period larger
%than 78 min is 4 times more likely to be the result of a CV with a
%companion that became substellar due to evolution, than of a CV with a
%companion that was born substellar.

\section{Conclusions}

We have used I-band spectroscopy in an attempt to detect
the mass donor of the cataclysmic variable GD~552.
Failure to detect the donor star puts strong limits on its mass,
ruling out a main sequence nature.

We have shown GD~552 data to be consistent with a model in which its
components are an ordinary white dwarf and a brown dwarf, at a moderate
inclination angle. Population synthesis calculations favour the idea
that the mass donor reached the stage of brown dwarf through evolution
instead of being born as such. This suggests that GD~552 is likely to be 
a post-period minimum CV, making it the ``period bouncer'' with the longest 
securely-determined orbital period known.

We confirm the value obtained by HH1990 for the radial velocity semiamplitude
of the white dwarf and determine for it a $T_{\rm eff}$ of $10900\pm400$ K
by fitting the spectra with a grid of white dwarf models. The temperature 
determined for the white dwarf in GD552 is consistent with the red edge of the
instability strip (11\,000 $< T_{\rm eff} <$ 12\,000 K; Bergeron et
al. 1995). From the 5.4\,h stretch of photometry presented, we find no
indication of the presence of pulsations. The lightcurve shows
significant varibility but not with the periodicities and amplitudes
characteristic of ZZ Cet stars.

\section*{Acknowledgments}

EU was supported by PPARC (UK) and Fundaci\'on Andes (Chile) under
the program Gemini PPARC-Andes throughout most of this research.
EU and RH-G were also supported by Universidad Cat\'olica del Norte's
Proyecto de Investigaci\'on en Docencia 220401-10602015.
TRM was supported under a PPARC SRF during some of the period over
which this work was undertaken. LM-R is supported by NWO VIDI grant
639.042.201 to P. J. Groot. BTG was supported by a PPARC Advanced
Fellowship.

The authors acknowledge the data analysis facilities provided by the
Starlink Project which is run by the University of Southampton on
behalf of PPARC. This research has made use of NASA's Astrophysics 
Data System Bibliographic Services. We acknowledge with thanks the
variable star observations from the AAVSO International
Database contributed by observers worldwide and used in this research.

The INT and the WHT are operated on the island of La Palma by the
Isaac Newton Group in the Spanish Observatorio del Roque de los Muchachos of
the Instituto de Astrofisica de Canarias 
This research was also partly based on observations made with the NASA/ESA Hubble 
Space Telescope, obtained from the data archive at the Space Telescope Science 
Institute. STScI is operated by the Association of Universities for Research in 
Astronomy, Inc. under NASA contract NAS 5-26555; these observations are
associated wirh program 9406.

\label{lastpage}
\end{document}